\documentclass[11pt,a4paper]{article}
\pdfoutput=1
\usepackage{jheppub}

\makeatletter
\gdef\@fpheader{Published in JHEP 09 (2026) 216}
\makeatother

\usepackage{etoolbox} 
\usepackage{graphicx} 
\usepackage{epstopdf} 
\usepackage{dcolumn} 
\usepackage{xcolor} 
\usepackage{amsmath,amssymb} 
\usepackage{hyperref} 
\usepackage[utf8]{inputenc} 
\usepackage[english]{babel} 
\usepackage{ifthen} 
\usepackage{blindtext} 
\usepackage{orcidlink} 
\usepackage{datetime2} 
\ifthenelse{\isundefined{\revtex}}{\usepackage[displaymath, mathlines]{lineno}}{} 

\graphicspath{{figures/}} 

\newboolean{articletitles}
\setboolean{articletitles}{true} 

\RequirePackage{xspace}
\RequirePackage{relsize}

\def\babar{\mbox{\slshape B\kern-0.1em{\smaller A}\kern-0.1em
    B\kern-0.1em{\smaller A\kern-0.2em R}}\xspace}

\def\Pgamma      {\ensuremath{\gamma}\xspace}

\def\Ppi         {\ensuremath{\pi}\xspace}

\def\Ppsi        {\ensuremath{\psi}\xspace}                 
                 
\mathchardef\PDelta="7101
\mathchardef\PXi="7104
\mathchardef\PLambda="7103
\mathchardef\PSigma="7106
\mathchardef\POmega="710A
\mathchardef\PUpsilon="7107
                 
\def\PB      {\ensuremath{B}\xspace}                 
                 
\def\PD      {\ensuremath{D}\xspace}

\def\PJ      {\ensuremath{J}\xspace}                 
\def\PK      {\ensuremath{K}\xspace}

\def\Pd      {\ensuremath{d}\xspace}

\def\Pq      {\ensuremath{q}\xspace}

\def\g      {\ensuremath{\Pgamma}\xspace}

\def\quark     {\ensuremath{\Pq}\xspace}
\def\quarkbar  {\ensuremath{\overline \quark}\xspace}
\def\qqbar     {\ensuremath{\quark\quarkbar}\xspace}

\def\dquark    {\ensuremath{\Pd}\xspace}

\def\pion  {\ensuremath{\Ppi}\xspace}
\def\piz   {\ensuremath{\pion^0}\xspace}

\def\kaon  {\ensuremath{\PK}\xspace}
\def\Kbar  {\kern 0.2em\overline{\kern -0.2em \PK}{}\xspace}

\def\Kz    {\ensuremath{\kaon^0}\xspace}
\def\Kzb   {\ensuremath{\Kbar^0}\xspace}
\def\KzKzb {\ensuremath{\Kz \kern -0.16em \Kzb}\xspace}
\def\Kp    {\ensuremath{\kaon^+}\xspace}
\def\Km    {\ensuremath{\kaon^-}\xspace}

\def\KpKm  {\ensuremath{\Kp \kern -0.16em \Km}\xspace}
\def\KS    {\ensuremath{\kaon^0_{\rm\scriptscriptstyle S}}\xspace} 
\def\KL    {\ensuremath{\kaon^0_{\rm\scriptscriptstyle L}}\xspace}

\def\D       {\ensuremath{\PD}\xspace}
\def\Dbar    {\kern 0.2em\overline{\kern -0.2em \PD}{}\xspace}

\def\Dz      {\ensuremath{\D^0}\xspace}
\def\Dzb     {\ensuremath{\Dbar^0}\xspace}
\def\DzDzb   {\ensuremath{\Dz {\kern -0.16em \Dzb}}\xspace}
\def\Dp      {\ensuremath{\D^+}\xspace}
\def\Dm      {\ensuremath{\D^-}\xspace}

\def\DpDm    {\ensuremath{\Dp {\kern -0.16em \Dm}}\xspace}

\def\B       {\ensuremath{\PB}\xspace}
\def\Bbar    {\ensuremath{\kern 0.18em\overline{\kern -0.18em \PB}{}}\xspace}

\def\BBbar   {\ensuremath{\B\Bbar}\xspace} 
\def\Bz      {\ensuremath{\B^0}\xspace}
\def\Bzb     {\ensuremath{\Bbar^0}\xspace}

\def\Bd      {\ensuremath{\B^0}\xspace}

\def\jpsi     {\ensuremath{{\PJ\mskip -3mu/\mskip -2mu\Ppsi\mskip 2mu}}\xspace}

\def\Y#1S{\ensuremath{\PUpsilon{(#1S)}}\xspace}

\def\Lbar {\ensuremath{\kern 0.1em\overline{\kern -0.1em\PLambda}}\xspace}

\def\to                 {\ensuremath{\rightarrow}\xspace}

\def\CP                {\ensuremath{C\!P}\xspace}

\newcommand{\dmd}{\ensuremath{\Delta m_{\dquark}}\xspace}

\newcommand{\ACP}{\ensuremath{{\cal A}^{\CP}}\xspace}

\newcommand{\tev}{\ensuremath{\mathrm{\,Te\kern -0.1em V}}\xspace}
\newcommand{\gev}{\ensuremath{\mathrm{\,Ge\kern -0.1em V}}\xspace}
\newcommand{\mev}{\ensuremath{\mathrm{\,Me\kern -0.1em V}}\xspace}
\newcommand{\kev}{\ensuremath{\mathrm{\,ke\kern -0.1em V}}\xspace}
\newcommand{\ev}{\ensuremath{\mathrm{\,e\kern -0.1em V}}\xspace}
\newcommand{\gevc}{\ensuremath{{\mathrm{\,Ge\kern -0.1em V\!/}c}}\xspace}
\newcommand{\mevc}{\ensuremath{{\mathrm{\,Me\kern -0.1em V\!/}c}}\xspace}
\newcommand{\gevcc}{\ensuremath{{\mathrm{\,Ge\kern -0.1em V\!/}c^2}}\xspace}
\newcommand{\gevgevcccc}{\ensuremath{{\mathrm{\,Ge\kern -0.1em V^2\!/}c^4}}\xspace}
\newcommand{\mevcc}{\ensuremath{{\mathrm{\,Me\kern -0.1em V\!/}c^2}}\xspace}

\def\gsim{{~\raise.15em\hbox{$>$}\kern-.85em
          \lower.35em\hbox{$\sim$}~}\xspace}
\def\lsim{{~\raise.15em\hbox{$<$}\kern-.85em
          \lower.35em\hbox{$\sim$}~}\xspace}

\def\evtgen   {\mbox{\textsc{EvtGen}}\xspace}
\def\pythia    {\mbox{\textsc{Pythia8}}\xspace}
\def\kkmc     {\mbox{\textsc{KKMC}}\xspace}
\def\geant     {\mbox{\textsc{Geant4}}\xspace}

\usepackage{multirow}
\usepackage{float} 
\usepackage{subcaption} 

\def\Pi         {\ensuremath{i}\xspace}
\def\CS         {\ensuremath{C^\prime}\xspace}
\newcommand{\dt}{\ensuremath{\Delta t}\xspace}
\newcommand{\dtres}{\ensuremath{\Delta t_{\rm res}}\xspace}

\newcommand{\sdt}{\ensuremath{\sigma_{\dt}}\xspace}
\renewcommand{\ACP}{\ensuremath{{\cal A}_{\CP}}\xspace}
\newcommand{\CCP}{\ensuremath{C}\xspace}
\newcommand{\SCP}{\ensuremath{S}\xspace}
\newcommand{\Mbc}{\ensuremath{M_{\mathrm{bc}}}\xspace}
\newcommand{\DeltaE}{\ensuremath{\Delta E}\xspace}
\newcommand{\pvec}[1]{\vec{#1}\mkern2mu\vphantom{#1}}
\newcommand{\brac}[1]{\left(#1\right)}
\def\pythiab    {\mbox{\textsc{Pythia6}}\xspace}
\def\geantb     {\mbox{\textsc{Geant3}}\xspace}

\begin{document}

\title{Measurement of time-dependent {\boldmath $\CP$} violation parameters in {\boldmath $\Bd\to\KS\piz\g$} decays at Belle and Belle~II
}
\collaboration{The Belle and Belle II Collaborations}
  \author{M.~Abumusabh\,\orcidlink{0009-0004-1031-5425},} 
  \author{I.~Adachi\,\orcidlink{0000-0003-2287-0173},} 
  \author{A.~Aggarwal\,\orcidlink{0000-0002-5623-3896},} 
  \author{Y.~Ahn\,\orcidlink{0000-0001-6820-0576},} 
  \author{H.~Aihara\,\orcidlink{0000-0002-1907-5964},} 
  \author{M.~Akdag\,\orcidlink{0009-0004-3728-1077},} 
  \author{N.~Akopov\,\orcidlink{0000-0002-4425-2096},} 
  \author{S.~Alghamdi\,\orcidlink{0000-0001-7609-112X},} 
  \author{M.~Alhakami\,\orcidlink{0000-0002-2234-8628},} 
  \author{N.~Althubiti\,\orcidlink{0000-0003-1513-0409},} 
% ...

\abstract{We perform a measurement of time-dependent \CP violation parameters in $\Bd \to \KS \piz \g$ decays using a dataset of approximately $772 \times 10^6$ and $521 \times 10^6$ $\Y4S$ decays collected by the Belle and Belle~II experiments, respectively.
The measured parameters for the combined dataset in the $K^{*0}(892)$ dominated region ($M_{\KS\piz} \in [0.8,1.0]\gevcc$) are $\SCP = 0.09 \pm 0.16 \pm 0.02$ and $\CCP = -0.09 \pm 0.08 \pm 0.04$.
For the non-$K^{*0}(892)$ region ($M_{\KS\piz} \in (1.0,1.8]\gevcc$), the corresponding values are $\SCP = -0.32 \pm 0.33 \pm 0.09$ and $\CCP = -0.07 \pm 0.17 \pm 0.08$.
The first quoted uncertainties are statistical, while the second ones are systematic.
These results are consistent with Standard Model predictions and more precise than previous measurements.}

\maketitle
\flushbottom

\section{Introduction}

The study of time-dependent \CP violation in the radiative decay $\B \to \KS \piz \g$ provides a unique probe for physics beyond the Standard Model (SM). The decay occurs predominantly through \( b \rightarrow s \gamma \) loop transitions, which makes it susceptible to contributions from heavy, virtual particles~\cite{Atwood:1997zr,Hewett:1998}.  As a result, this decay channel can explore physics at energy scales significantly larger than those directly accessible in current collider experiments.
 
The time-dependent \CP asymmetry arises due to interference between decay amplitudes with and without $\Bz$--$\Bzb$ mixing, where a nonzero phase appearing in the quark-mixing matrix causes differing decay rates for $\Bz$ and $\Bzb$ over time.
For coherent \BBbar pair production at the \Y4S, in which one of the \B mesons (\(B_{\text{sig}}\)) decays to a final state $f_{\CP}\gamma$, where $f_{\CP}$ is a \CP eigenstate, and the other \B meson (\(B_{\text{tag}}\)) decays to a flavour-specific final state, the \CP asymmetry is given as
\begin{equation}
    \ACP(\Delta t) = \frac{\Gamma(B_{\text{tag}=\Bz}(\Delta t) \to f_{\CP}\gamma) - \Gamma(B_{\text{tag}=\Bzb}(\Delta t) \to f_{\CP}\gamma)}{\Gamma(B_{\text{tag}=\Bz}(\Delta t) \to f_{\CP}\gamma)  + \Gamma(B_{\text{tag}=\Bzb}(\Delta t) \to f_{\CP}\gamma)},
\end{equation}
where $\Gamma(B_{{\rm tag}=\Bz/\Bzb})$ is the decay rate of $B_{\text{sig}}$ to $f_{\CP}\gamma$ when $B_{\text{tag}}$ has been identified as a $\Bz/\Bzb$ meson, and \( \Delta t \) is the proper time difference between the decays of \( B_{\text{sig}} \) and \( B_{\text{tag}} \). This asymmetry can be parameterised as
\begin{equation}
    \ACP(\Delta t) = \SCP \sin(\Delta m_d \Delta t) - \CCP \cos(\Delta m_d \Delta t),
    \label{cp eqn}
\end{equation}
where \( \Delta m_d \) is the mass difference between the two \( B^0 \) mass eigenstates, and \( \SCP \) and \( \CCP \) are the mixing-induced and direct \CP-violating parameters, respectively.

In the SM, the photon emitted in \( b \rightarrow s \gamma \) processes is predominantly left-handed due to the chiral structure of the weak interaction. The presence of a polarised photon in the final state suppresses the interference, leading to small \CP-violating effects. In particular, the SM predicts that \SCP should be suppressed by a factor of $m_{s}/m_{b}$ due to helicity suppression, with deviations expected at the level of a few percent~\cite{PhysRevD.75.054004}. Here, $m_s$ ($m_b$) is the mass of the strange (bottom) quark. The \SCP value is calculated to be $(-2.3\pm 1.6)\%$ for the resonant $B^0 \to K^{*0}(892)\g$ channel, which can, however, be enhanced to ${\cal O}(10\%)$ in nonresonant $\B \to \KS \piz \g$ decays through long-distance effects, such as charm loop contributions~\cite{PhysRevD.71.011504, PhysRevD.73.014013}. The \CCP value is expected to be smaller than 1\%~\cite{Kagan:1998bh}; however, this estimate may not be robust due to potentially large uncertainties in predicting strong phases~\cite{PhysRevD.75.054004}.

Any significant contribution from the right-handed photon can increase the \SCP value beyond the previously mentioned predictions, indicating evidence for new physics, such as supersymmetry, extended Higgs sectors, or models with vector-like fermions~\cite{np1, np2, Kou:2010kn}. Therefore, a precise measurement of the time-dependent \CP asymmetry in  $\B \to \KS \piz \g$ provides a stringent test of the SM and a powerful probe for new sources of \CP violation from physics beyond the SM. Moreover, the different mechanisms at play in the theoretical predictions~\cite{PhysRevD.75.054004, PhysRevD.71.011504, PhysRevD.73.014013} for the $B^0 \to K^{*0}(892)\g$ and nonresonant decays motivate separate investigation of these channels.

The previous measurements from Belle, BaBar, and Belle~II are summarised in Table~\ref{tab:previous_results}. While these results are consistent with SM predictions, they exhibit significant uncertainties. The objective of our study is to enhance the measurement precision by utilizing a larger dataset and advanced algorithms. In particular, we have implemented an improved, Graph Neural Network (GNN)-based flavour Tagger~\cite{PhysRevD.110.012001} across both the Belle and Belle II datasets, along with an improved \KS selection algorithm specifically for the Belle dataset. We have also incorporated information from events with poor time resolution in a time-integrated fit, which contributes to reducing the uncertainty on \CCP.
\begin{table}[h]
\centering
    \caption{Summary of earlier measurements of \SCP and \CCP for the $\Bz \to K^{*0}(892) \g$ and $\Bz \to \KS \piz \g$ channels excluding the resonant $K^{*0}(892)$ contribution. The first quoted uncertainties are statistical, while the second ones are systematic.}
    \begin{tabular}{p{2cm}l c l l}
        Channel & Experiment & $N_{\Upsilon(4S)}$ & \multicolumn{1}{c}{$\SCP$} & \multicolumn{1}{c}{$\CCP$} \\
        \hline\hline
        \multirow{3}{=}{$K^{*0}(892) \g$} & Belle~\cite{Belle:2006pxp} & $535\times10^6$ & $-0.32^{\,+\,0.36}_{\,-\,0.33} \pm 0.05$ & $\phantom{-}0.20 \pm 0.24 \pm 0.05$ \\
        & BaBar~\cite{BaBar:2008okc} & $467\times10^6$ & $-0.03 \pm 0.29 \pm 0.03$ & $-0.14 \pm 0.16 \pm 0.03$ \\
        & Belle~II~\cite{PhysRevLett.134.011802} & $388\times10^6$ & $\phantom{-}0.00^{\,+\,0.27}_{\,-\,0.26} \pm 0.03$ & $\phantom{-}0.10 \pm 0.13 \pm 0.04$ \\
        \hline
        \multirow{3}{=}{$\KS \piz \g$, excluding $K^{*0}(892) \g$} & Belle~\cite{Belle:2006pxp} & $535\times10^6$ & $\phantom{-}0.50 \pm 0.61 \pm 0.29$ & $\phantom{-}0.20 \pm 0.37 \pm 0.13$ \\
        & BaBar~\cite{BaBar:2008okc} & $467\times10^6$ & $-0.78 \pm 0.59 \pm 0.09$ & $-0.36 \pm 0.33 \pm 0.04$ \\
        & Belle~II~\cite{PhysRevLett.134.011802} & $388\times10^6$ & $\phantom{-}0.04^{\,+\,0.45}_{\,-\,0.44} \pm 0.23$ & $-0.06 \pm 0.25 \pm 0.09$ \\
    \end{tabular}
    \label{tab:previous_results}
\end{table}

\section{Detectors and data samples}

The Belle and Belle II detectors both have a cylindrical geometry whose symmetry ($z$) axis is nearly aligned with the electron beam direction at the interaction point (IP). The polar angle $\theta$ is defined relative to the $z$ axis.

The Belle detector~\cite{Abashian2002117} was located at the KEKB accelerator~\cite{Kurokawa:2001nw}, which collided electrons and positrons with beam energies of 8.0\,GeV and 3.5\,GeV, respectively. It recorded data from 1999 to 2010. Belle was a large-solid-angle magnetic spectrometer composed of a silicon vertex detector (SVD), a central drift chamber (CDC), an array of aerogel threshold Cherenkov counters (ACC), a barrel-like arrangement of time-of-flight (TOF) scintillation counters, and an electromagnetic calorimeter (ECL) comprised of CsI(Tl) crystals, all located inside a superconducting solenoid coil that provided a magnetic field of 1.5\,T. The SVD and CDC were used to reconstruct charged particle tracks and vertices, while the ACC and TOF, along with specific ionization ($dE/dx$) measurements from the CDC, were used for charged particle identification (PID) purposes, and photons were reconstructed from calorimetric clusters in the ECL. An iron flux-return yoke, placed outside the coil, was instrumented with resistive-plate chambers to detect $\KL$ mesons and muons.

The Belle II detector~\cite{Abe:2010gxa} represents a significant upgrade from Belle, and is currently operational at the SuperKEKB accelerator~\cite{Akai:2018mbz}, which collides electrons and positrons with beam energies of 7.0\,GeV and 4.0\,GeV, respectively.
Belle II features a two-layer pixel detector (PXD) that is surrounded by a four-layer SVD~\cite{Belle-IISVD:2022upf} and a 56-layer CDC. These subdetectors help reconstruct the tracks of charged particles; the SVD and CDC additionally provide $dE/dx$ information for PID purposes. Enclosing the CDC are a time-of-propagation counter (TOP)~\cite{ATMACAN2025170627} located in the central region and an aerogel-based ring-imaging Cherenkov counter (ARICH) situated in the forward region. Together, these subdetectors constitute the primary PID system. Surrounding both the TOP and ARICH is the ECL, made from the same CsI(Tl) crystals as Belle but with faster electronics, which measures energy and time for photons and electrons. 
Outside the ECL is a superconducting solenoid magnet that provides a magnetic field of 1.5\,T, parallel to the $z$ axis.
The flux return of the magnet is equipped with resistive-plate chambers and plastic scintillator modules for the detection of muons, $\KL$ mesons, and neutrons.

The data samples used in our study were collected at the $\Upsilon(4S)$ resonance, which decays predominantly into $B\overline{B}$ pairs. The Belle dataset corresponds to an integrated luminosity of $711\,{\rm fb}^{-1}$, containing approximately $772 \times 10^6$ $\Y4S$ decays. Similarly, the Belle II dataset used corresponds to an integrated luminosity of $493\,{\rm fb}^{-1}$, containing approximately $521 \times 10^6$ $\Y4S$ decays. About 81\% of the latter sample was collected with both the PXD and SVD operational, while the remaining 19\% was collected with the PXD off.

The analysis strategy and fitting procedure are developed using simulated Monte Carlo (MC) samples. Data control samples and sidebands are then used to validate the procedures before examining the signal region in the data, which is accessed only after the analysis procedures have been finalized. MC simulation samples for the process $e^{+}e^{-} \to \Upsilon(4S) \to B\overline{B}$ are generated using the \evtgen~\cite{Lange:2001uf} package, designed for $B$ meson decays. The $e^{+}e^{-} \to q\overline{q}$ ($q \in [u, d, s, c]$) continuum events are generated using the \kkmc~\cite{Jadach:1999vf} package. Both packages are interfaced with \pythiab~\cite{pythia6} (\pythia~\cite{Sjostrand:2014zea}) for Belle (Belle~II).
The detector response is modeled with the \geantb~\cite{geant3} framework for Belle and \geant~\cite{Agostinelli:2002hh} for Belle~II. We use MC samples from generic $e^{+}e^{-}$ collisions, i.e., combining $\Bz\Bzb$, $B^{+}B^{-}$, and \qqbar events, as well as specific samples of \BBbar events, where one of the $B$ mesons decays into a specified mode of interest, for example, the signal. The Belle II analysis software framework~\cite{Kuhr:2018lps, Gelb:2018} is used to process both the simulated and real data samples.

\section{Reconstruction and event selection}

The reconstruction of $\Bz \to \KS \piz \g$ candidates is performed with a bottom-up approach, where $\KS$ candidates are first reconstructed from tracks in the PXD (Belle~II), SVD, and CDC, while $\piz$ and $\g$ candidates are reconstructed from clusters in the ECL. The remaining tracks and clusters are assigned to the $B_{\rm tag}$ candidates.

We reconstruct \KS using two oppositely charged tracks, assumed to be pions and constrained to come from a common vertex, and that their invariant mass lies within $\pm 30~\mevcc$ of the known \KS mass~\cite{PDG2022}. To suppress misreconstructed \KS candidates, two multivariate (MVA) classifiers are used. The first one selects \KS--like candidates by primarily using the kinematic properties of the \KS candidate and its decay pions, together with its flight length. The second MVA is designed to reduce the contamination from $\Lambda^0 \to p^+ \pi^-$ decays by using the proton PID likelihood of each candidate decay pion, and the reconstructed $\Lambda^0$ mass, obtained by assigning the proton mass hypothesis to one of the pion candidates. In Belle, these classifiers~\cite{PhysRevD.97.092003} are based on a NeuroBayes Neural Network~\cite{FEINDT2006190} package while in Belle~II, LightGBM~\cite{LightGBM} is employed. 

We reconstruct \piz candidates from two photon candidates with invariant mass satisfying $m_{\g\g}\in [90,160]~\mevcc$, with further constraints on their shower shape in the ECL and the opening angle between them. For Belle~II, we also require the ECL cluster time for each photon candidate to be within $\pm 200\,{\rm ns}$ of the beam-crossing time. A mass-constrained fit is applied to improve the momentum resolution of \piz candidates. To reject misreconstructed \piz candidates, separate MVA classifiers based on XGBoost~\cite{ChenGuestrin2016} are employed for Belle and Belle~II utilizing features related to the kinematics of the \piz candidates and the shower shapes of their decay products. The key discriminating features are the $\chi^2$ probability of the mass-constrained fit and the opening angle between the \piz decay products.

Prompt photon candidates are selected from high-energy ECL clusters with energies satisfying $E^{*} > 1.4 \gev$. The asterisk ($*$) denotes quantities calculated in the $e^{+}e^{-}$ center-of-mass frame. To ensure the cluster shape is consistent with an electromagnetic shower, we require that the ratio $E_9$/$E_{21}$ exceeds 0.9, where $E_9$ and $E_{21}$ are the energies deposited in a $3\times3$ array of crystals and a $5\times5$ array excluding the four corners, respectively, centred on the crystal with the highest energy. Furthermore, for Belle~II, we require the ECL cluster time for each photon candidate to be within $\pm 200\,{\rm  ns}$ of the beam-crossing time. To suppress backgrounds from $\pi^0 \to \gamma \gamma$ and $\eta \to \gamma \gamma$ decays, separate MVA classifiers based on XGBoost are employed for Belle and Belle~II~\cite{PhysRevD.111.L071103}. These classifiers combine the kinematic information of the prompt photon candidate with other clusters from the rest of the event to identify potential $\piz$ or $\eta$ candidates. If such candidates are found, then the prompt photon candidate is rejected. The most discriminating features are the reconstructed $\piz/\eta$ mass and the $\chi^2$ probability of the corresponding mass-constrained fit.

The $B_{\rm sig}$ candidates are reconstructed by combining the selected \KS, \piz, and prompt photon candidates. A vertex fit is performed to the entire decay chain with the $B_{\rm sig}$ candidates constrained to originate from the IP~\cite{Krohn:2019dlq}. An additional $B_{\rm sig}$ vertex fit is performed with the $\piz$ mass constraint, which is solely aimed at improving the $\piz$ four-momentum resolution by using the $B_{\rm sig}$ vertex as the $\piz$ production point.

Two kinematic variables are used to identify $B_{\rm sig}$ candidates: the beam-constrained mass \( \Mbc = (1/c^2)\sqrt{s/4 - (|\pvec{p}^*_B|c)^2} \) and the energy difference \( \DeltaE = E^*_B - \sqrt{s}/2 \), where \( \sqrt{s}/2 \) is the beam energy, and \( E^*_B \) and \( \pvec{p}^*_B \) are the energy and momentum of the reconstructed $B_{\rm sig}$ candidate, respectively. Owing to the presence of a high-energy photon in the final state, the \DeltaE distribution is asymmetric, with a long tail on the negative side arising primarily due to shower leakage in the ECL. This effect also causes a correlation between \Mbc and \DeltaE~\cite{Bevan:2014iga}, which can be mitigated by redefining \Mbc as follows:
\begin{align}
    E_{\rm res} &= \sqrt{s}/2 - E^*_{\KS}, \nonumber\\
    E_{\piz}' &= \frac{E_{\rm res}}{E^*_{\piz} + E^*_{\g}} \times E^*_{\piz},\ E_{\g}' = \frac{E_{\rm res}}{E^*_{\piz} + E^*_{\g}} \times E^*_{\g}, \nonumber\\
    {\pvec{p}_{\piz}'} &= \sqrt{(E_{\piz}'/c)^2 - (M_{\piz}c)^2}\times \frac{\pvec{p}^*_{\piz}}{|\pvec{p}^*_{\piz}|},\ {\pvec{p}_{\g}'} = (E_{\g}'/c)\times \frac{\pvec{p}^*_{\g}}{|\pvec{p}^*_{\g}|}, \nonumber\\
    \Mbc &= (1/c^2)\sqrt{s/4 - (|\vec{p}_{\KS} + \pvec{p}_{\piz}' + \pvec{p}_{\g}'|c)^2}.
    \label{eqn:mbc}
\end{align}
By rescaling the \piz and photon energies using the precisely measured kinematics of the \KS and the $e^+e^-$ collision system, we minimize the dependence of \Mbc on the measured energies of the \piz and photon candidates, which are prone to biases from shower leakage. This rescaling significantly reduces the correlation between \Mbc and \DeltaE in the background events without affecting the signal, and improves the signal \Mbc resolution by 3--7\% depending on the dataset. Candidates are required to satisfy \( \Mbc > 5.23~\gevcc \) and \( -0.3~\gev < \DeltaE < 0.2~\gev \). The invariant mass of the \( \KS \piz \) system, $M_{\KS\piz}$, is required to lie within the range $[0.8,1.8] \gevcc$. The dataset is divided into two regions: $M_{\KS\piz} \in [0.8,1.0] \gevcc$, denoted as MR1, and $M_{\KS\piz} \in (1.0,1.8] \gevcc$, denoted as MR2. The \( K^{*0}(892)\) resonance is the primary contributor in MR1, while MR2 is dominated by contributions from the nonresonant $\KS\piz\gamma$ channel and higher kaonic resonances such as \( K^{*0}(1410)\), \( K^{*0}(1430)\), \( K^{*0}(1680)\).

We reconstruct the $B_{\rm tag}$ vertex using well-reconstructed tracks not associated with the $B_{\rm sig}$ candidate. A further constraint is applied to ensure that the vertex lies in a tube originating from the IP, along the $B_{\rm tag}$ flight direction calculated using the $B_{\rm sig}$ momentum~\cite{iptube}. We retain events where both $B_{\rm sig}$ and $B_{\rm tag}$ vertices are successfully reconstructed. A GNN-based flavour tagging (FT) algorithm \cite{PhysRevD.110.012001} is employed to determine the flavour of $B_{\rm tag}$ at the time of its decay. This algorithm outputs a flavour tag \( q \) (where \( q = +1 \) for \( B^0 \) and \(-1 \) for \( \Bzb \)) and a quality factor \( r \) that indicates the confidence level of the flavour assignment.

The data include contributions from backgrounds associated with continuum events and \BBbar decays other than signal. To suppress the continuum background, we train MVA classifiers based on XGBoost, using event shape variables that leverage the topological differences between \qqbar events, which exhibit a boosted jet-like topology, and \BBbar events, characterised by a  nearly isotropic distribution. Separate classifiers are trained for Belle and Belle~II, as well as for MR1 and MR2, to account for differences in detector performance and background composition. The most discriminating feature in these classifiers is the angle between the thrust axes of the $B_{\rm sig}$ candidate and the rest of the event. The selection criteria for all MVA classifiers are determined by maximizing a figure-of-merit \( N_{\rm S}/\sqrt{N_{\rm S}+N_{\rm B}} \), where \( N_{\rm S} \) and \( N_{\rm B} \) are the numbers of signal and background events, respectively, expected in the signal region that is defined by $\Mbc > 5.27\gevcc$ and $-0.2 \gev < \DeltaE < 0.1 \gev$.
The asymmetric range for $\DeltaE$ is considered to account for the shower leakage in the ECL.  

The output of the continuum MVA classifier, after applying the selection criteria, is used as a variable in the final fit. This MVA output exhibits a strong peak near 1.0 for the signal, making it challenging to model. We therefore transform it using the following relation to obtain a variable that can be more readily modeled with analytical functions:
\begin{equation}
    \CS = \ln\left(\frac{C_{\mathrm{out}} - C_{\mathrm{out}}^{\mathrm{min}}}{C_{\mathrm{out}}^{\mathrm{max}} - C_{\mathrm{out}}}\right),
\end{equation}
where \( C_{\mathrm{out}} \) is the original MVA output, and \( C^{\mathrm{min}}_{\mathrm{out}} \) and \( C^{\mathrm{max}}_{\mathrm{out}} \) are the minimum and maximum values of the output, respectively.

After applying all selection criteria, 10--12\% of events have multiple signal candidates, depending on the dataset. Since these ambiguities are primarily due to misreconstructed \piz candidates, the candidate with the highest \piz MVA output is retained in such instances; for the remaining events with multiple candidates, a candidate is arbitrarily chosen. We find that this approach selects the correctly reconstructed signal decay in events with multiple candidates approximately 75\% of the time based on simulation studies.

The overall reconstruction and selection efficiency for signal events is estimated with simulated datasets. For Belle, the efficiency is approximately 19\% in MR1 and 13\% in MR2. For Belle~II, the corresponding values are higher, around 22\% in MR1 and 14\% in MR2; the increase is due to improved detector performance and reconstruction algorithms.

\section{Likelihood fit and results}

The distribution of the proper decay time difference \( \Delta t \) between $B_{\rm sig}$ and $B_{\rm tag}$ encodes the \CP asymmetries [Eq.~(\ref{cp eqn})]. This variable is calculated using the formula $\Delta t = \Delta \ell/\beta \gamma c$, where \( \Delta \ell \) is the distance between the two reconstructed vertices along the boost direction, and \( \beta \gamma \), the Lorentz boost of the \( \Upsilon(4S) \), equals $0.425$ ($0.284$) for Belle (Belle~II).

Based on the quality of \dt reconstruction, events are categorised into two groups: well- and poorly-reconstructed \dt events. For an event to be classified as well-reconstructed, both the \KS decay products must register at least one hit in the SVD layers for Belle, and in either the PXD or SVD layers for Belle~II, and the uncertainty in \dt must be less than $3.0\,{\rm ps}$ for both Belle and Belle~II. Such events account for approximately 50\% (75\%) of the total events selected in the Belle (Belle~II) dataset. The larger fraction in Belle~II arises from the larger volume of the vertex detector (PXD and SVD) and the smaller Lorentz boost in comparison to Belle, which enhance the likelihood of the \KS decaying within the vertex detector.

The well-reconstructed \dt events are used for time-dependent (TD) fits to extract both \SCP and \CCP parameters, while the poorly-reconstructed \dt events are used for time-integrated (TI) fits to extract only the \CCP parameter. To extract \SCP and \CCP, we perform a simultaneous extended unbinned maximum-likelihood fit to the TD and TI datasets in the two $B_{\rm tag}$ flavours. The TD fit involves \Mbc, \DeltaE, \CS, and \dt, taking per-event dependence on the \dt uncertainty ($\sigma_{\dt}$) and flavour-tagging quality factor $r$ into account. The TI fit uses \Mbc, \DeltaE, and \CS, considering per-event dependence on $r$. We use the same fit strategy for the Belle and Belle~II datasets in both MR1 and MR2. The fit model includes three components: signal (both correctly and partially reconstructed events with a well-reconstructed \KS), \BBbar, and \qqbar backgrounds. The probability density functions (PDFs) for all components are modelled with simulated events and validated with control channels.

Based on MC simulation studies, we find \Mbc, \DeltaE, and $\CS$ variables have negligible correlation with \dt, allowing us to factorise the full fit PDFs into the two parts: the signal extraction PDF and the \CP asymmetry PDF. Being a function of \Mbc, \DeltaE, and $\CS$, the first PDF aims to discriminate between various components and obtain the per-event signal probability. On the other hand, the second PDF models the \dt distribution with per-event dependence on $\sigma_{\Delta t}$ and $r$ to determine \SCP and \CCP. The full PDFs for TD and TI datasets are given by:
\begin{align}
    P_{\rm{full}}^{\rm TD}&(\Mbc, \DeltaE, \CS, \dt \mid \sigma_{\Delta t}, q, r) = 
    \sum_{i=1}^{3} N_{i}^{q}(q, r) f_{i}^{\rm TD} P_i^{\rm SE}(\Mbc, \DeltaE, \CS)  P_{i}^{\rm CPV}(\dt \mid \sigma_{\Delta t}, q, r)
\end{align}
and
\begin{align}
    P_{\rm{full}}^{\rm TI}&(\Mbc, \DeltaE, \CS \mid q, r) = 
    \sum_{i=1}^{3} N_{i}^{q}(q, r)(1-f_{i}^{\rm TD}) P_i^{\rm SE}(\Mbc, \DeltaE, \CS).
    \label{eqn:ti pdf}
\end{align}
Here, $i \in [1,3]$ represents the signal, \BBbar, and \qqbar components, respectively; $N_{i}^{q}(q, r)$ are the yields of components for the two $B_{\rm tag}$ flavours, expressed in terms of the direct \CP-violation parameter \CCP for signal and \BBbar background as described later; $f_{i}^{\rm TD}$ are the fractions of events in the TD dataset, obtained from simulation and calibrated with sidebands and control channels. Lastly, $P_{i}^{\rm SE}$ are the signal extraction PDFs, and $P_{i}^{\rm CPV}$ are the \CP asymmetry PDFs for the three components.

\subsection{Signal extraction PDF}

We model the shape in \Mbc for correctly reconstructed signal using a Crystal Ball function~\cite{Gaiser:Phd} with a Gaussian core and a polynomial tail on one side~\cite{Skwarnicki:1986xj}, and its \DeltaE shape using a Johnson's function~\cite{Johnson1949}. The correlation between \Mbc and \DeltaE can range from 5\% to 20\% depending on the dataset and is modelled in the fit by parameterising the width and tail parameters of the \Mbc PDF as functions of \DeltaE. The $\CS$ distribution for signal is modelled with a Johnson's function. The signal extraction PDF for signal is then the \Mbc--\DeltaE PDF multiplied by the $\CS$ PDF. For the partially reconstructed signal with a well-reconstructed \KS, known as self cross-feed (SCF), we use a Crystal Ball function for \Mbc, and an asymmetric Gaussian function for \DeltaE and $\CS$. The signal extraction PDF for the SCF component is the product of the individual \Mbc, \DeltaE, and $\CS$ PDFs. The final signal extraction PDF for the signal component ($P_{1}^{\rm SE}$) is the sum of the correctly reconstructed and SCF signal PDFs, weighted by their fractions expected from MC simulation.

The \BBbar background exhibits a peaking structure in \Mbc near the known $B$ meson mass~\cite{PDG2022}; however, such a structure is absent in \DeltaE. To address the residual correlation between \Mbc and \DeltaE for this background, we model its joint \Mbc--\DeltaE PDF using a two-dimensional Kernel Density Estimator (KDE)~\cite{CRANMER2001198}. The $\CS$ distribution is represented by an asymmetric Gaussian function.
The signal extraction PDF for the \BBbar background ($P_{2}^{\rm SE}$) is then the \Mbc--\DeltaE KDE PDF multiplied with the $\CS$ PDF.

We model the \qqbar background \Mbc shape using an ARGUS function~\cite{Albrecht:1989rj}, and its \DeltaE shape using a first-order polynomial. The \CS distribution for the \qqbar background is modelled with a Johnson's function.
Due to the negligible correlation among the three fit variables, we define the signal extraction PDF for the \qqbar background ($P_{3}^{\rm{SE}}$) as the product of the individual PDFs for \Mbc, \DeltaE, and $\CS$.

The final signal extraction PDF is the sum of signal, \BBbar, and \qqbar background signal extraction PDFs, weighted by their respective yields, which are floated in the fit:
\begin{equation}
    P^{\rm{SE}}_{\rm tot} = \sum_{i=1}^{3} N_{i} P_i^{\rm{SE}}(\Mbc, \DeltaE, \CS).
\end{equation}
The parameters describing the means of the \Mbc and \DeltaE functions for the signal PDF are floated in the fit. Similarly, the \Mbc and \DeltaE PDF shape parameters, along with the mean and width of the \CS PDF for the \qqbar component are floated in the  fit.

To validate the signal extraction fit strategy in the data and to determine potential data--MC correction factors for the fixed parameters, we use the $B^0 \to K^{*0} [K^+ \pi^-] \gamma$ control channel, following the same fit strategy as the signal channel. This channel is chosen because of its relatively large sample size and clean signature. It mimics the signal channel due to the presence of a high-energy photon in the final state. The means of the signal and \BBbar \CS PDFs, and the widths of the signal \Mbc, \DeltaE, and $\CS$ PDFs are floated in both data and simulation fits of the control channel to take into account possible data--MC differences. We extract as correction factors differences in the means and ratios in the widths, which  are of the order of 1--3\% for the signal and around 10\% for the \BBbar background. These correction factors are then applied to the full fit in the signal channel.

\begin{figure}[!h]
    \centering
    \begin{subfigure}{.33\textwidth}
        \centering
        \includegraphics[width=.9\linewidth]{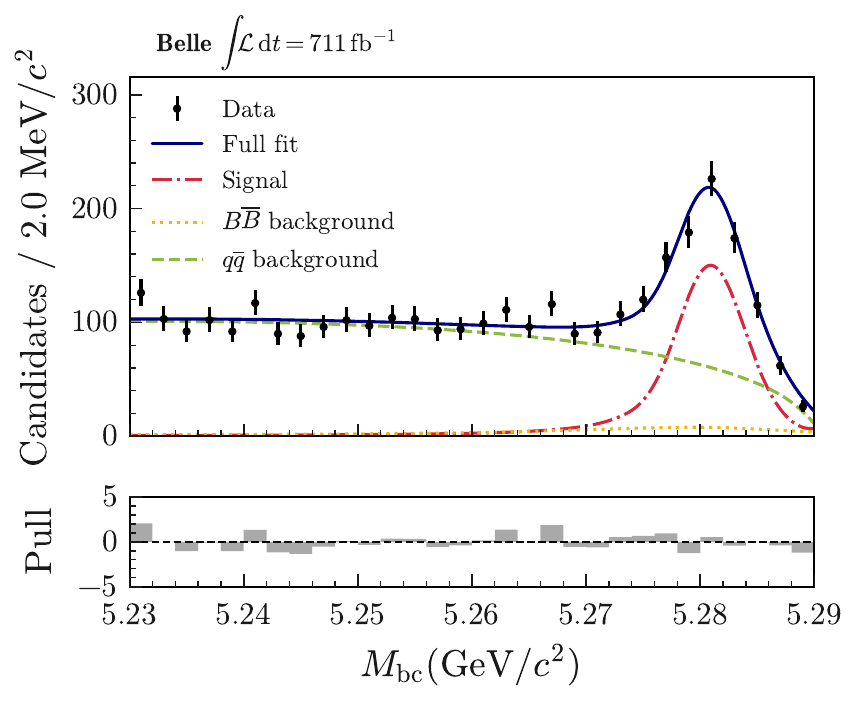}
    \end{subfigure}%
    \begin{subfigure}{.33\textwidth}
        \centering
        \includegraphics[width=.9\linewidth]{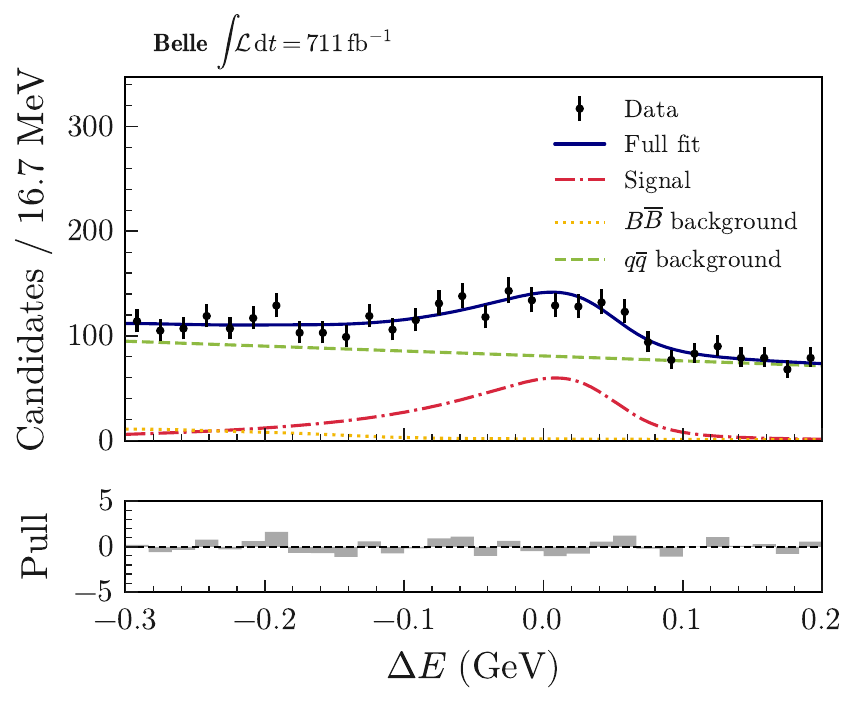}
    \end{subfigure}%
    \begin{subfigure}{.33\textwidth}
        \centering
        \includegraphics[width=.9\linewidth]{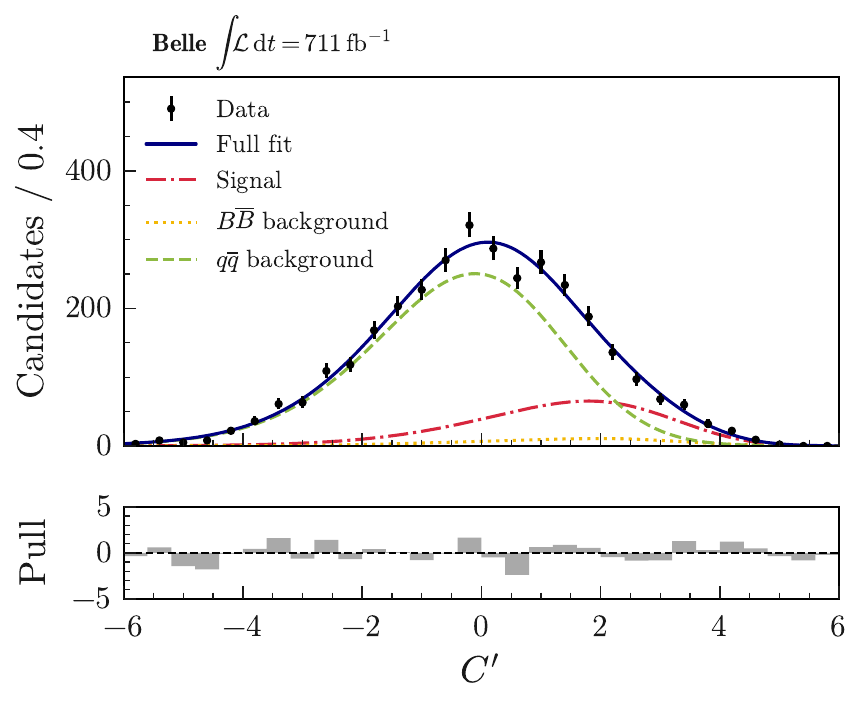}
    \end{subfigure}%
    \newline
    \begin{subfigure}{.33\textwidth}
        \centering
        \includegraphics[width=.9\linewidth]{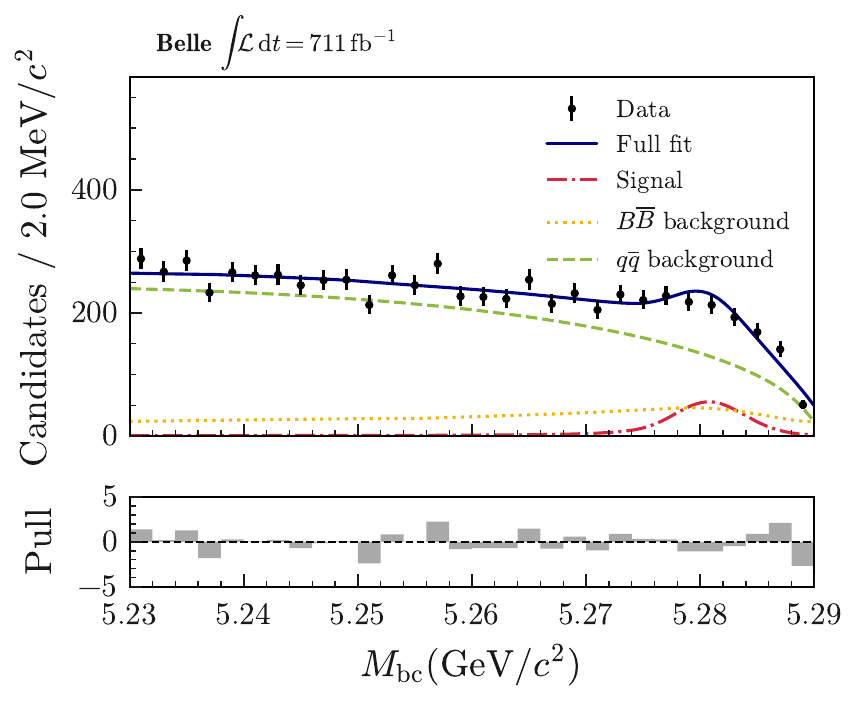}
    \end{subfigure}%
      \begin{subfigure}{.33\textwidth}
        \centering
        \includegraphics[width=.9\linewidth]{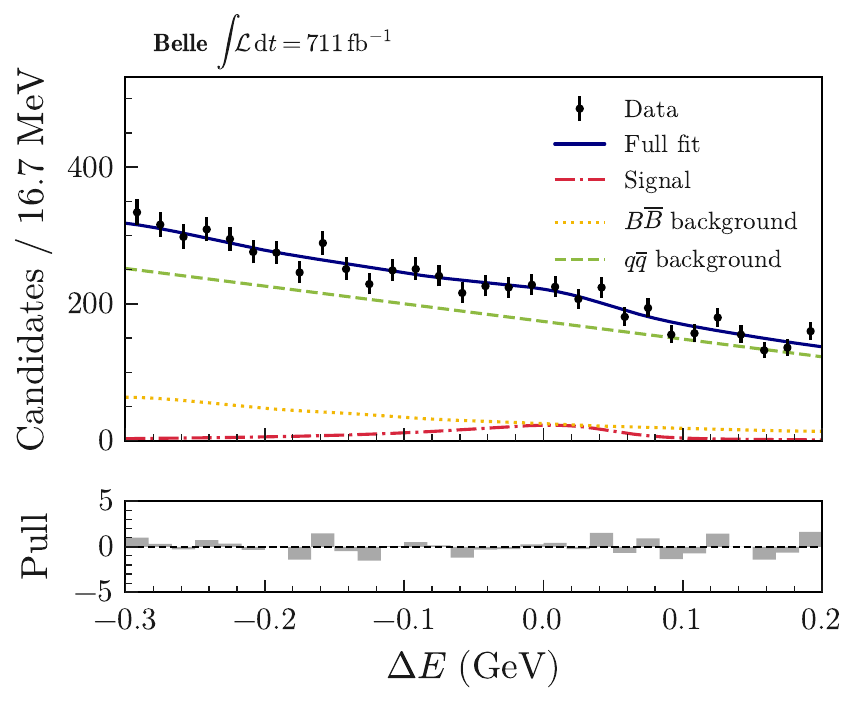}
      \end{subfigure}%
      \begin{subfigure}{.33\textwidth}
        \centering
        \includegraphics[width=.9\linewidth]{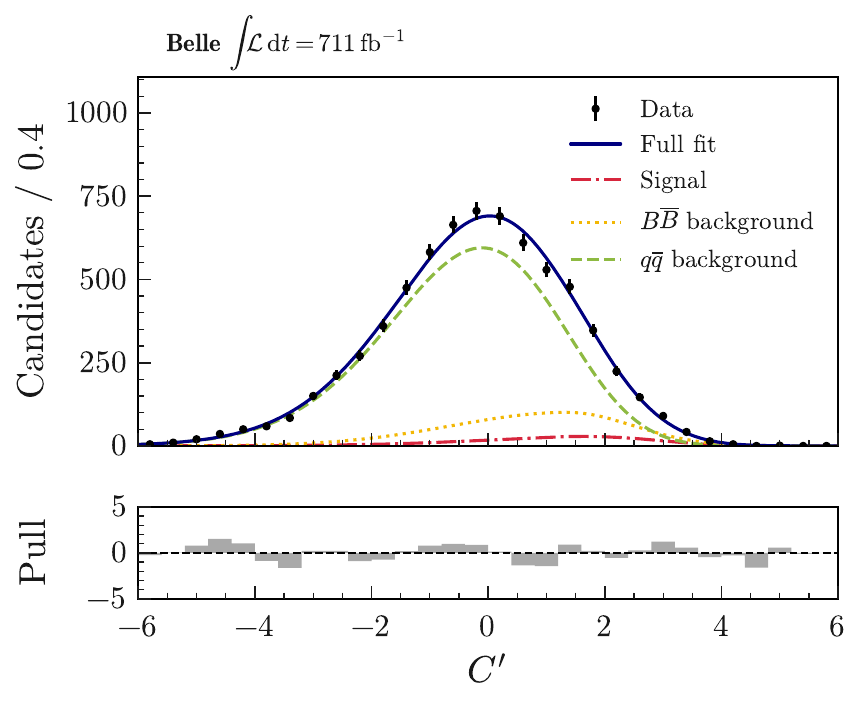}
      \end{subfigure}%
    \caption{\Mbc (left), \DeltaE (center), and $\CS$ (right) fit projections for MR1 (top) and MR2 (bottom) in Belle. Lower panels show the pulls, i.e., the differences between the data and fit results divided by the statistical uncertainty on the data.}
    \label{fig:se belle}
\end{figure}

\begin{figure}[!h]
    \centering
    \begin{subfigure}{.33\textwidth}
        \centering
        \includegraphics[width=.9\linewidth]{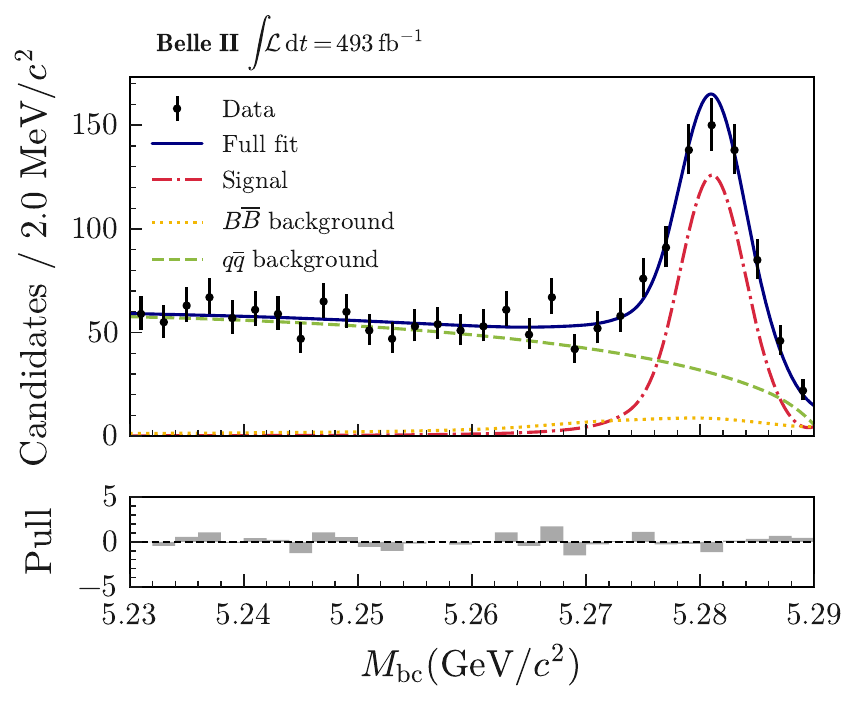}
      \end{subfigure}%
      \begin{subfigure}{.33\textwidth}
        \centering
        \includegraphics[width=.9\linewidth]{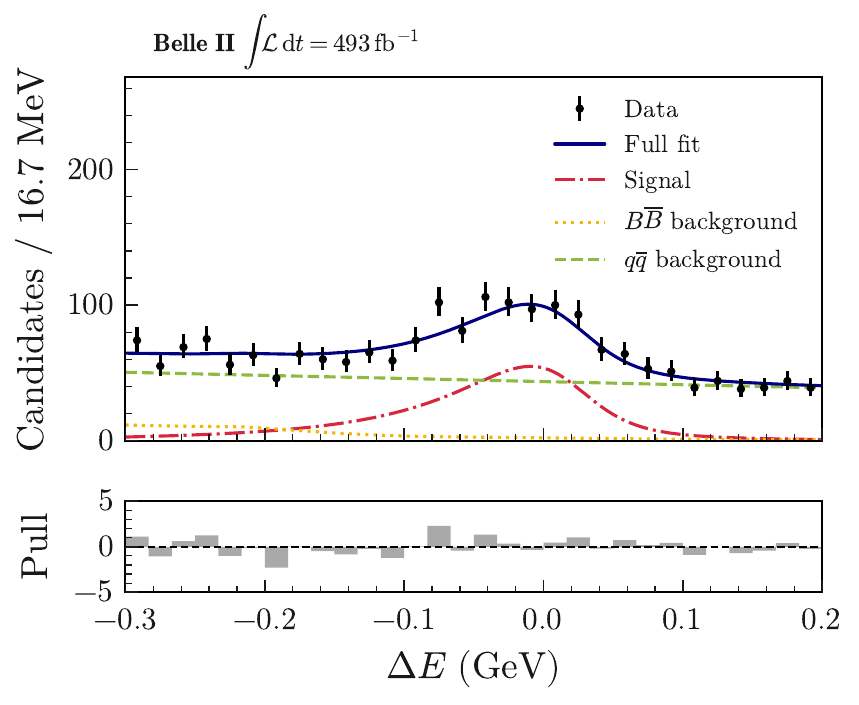}
      \end{subfigure}%
      \begin{subfigure}{.33\textwidth}
        \centering
        \includegraphics[width=.9\linewidth]{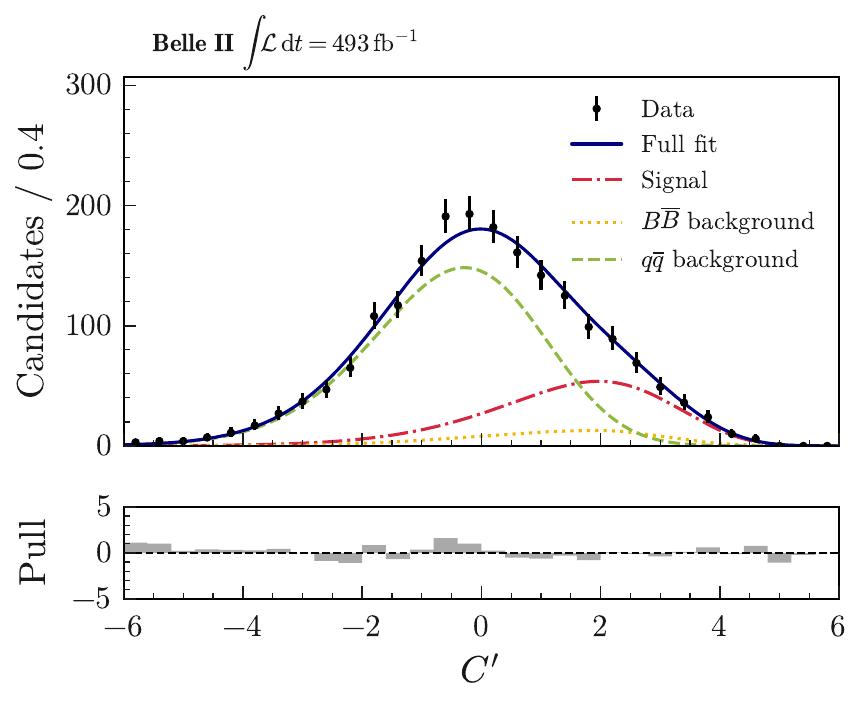}
      \end{subfigure}%
      \newline
    \begin{subfigure}{.33\textwidth}
        \centering
        \includegraphics[width=.9\linewidth]{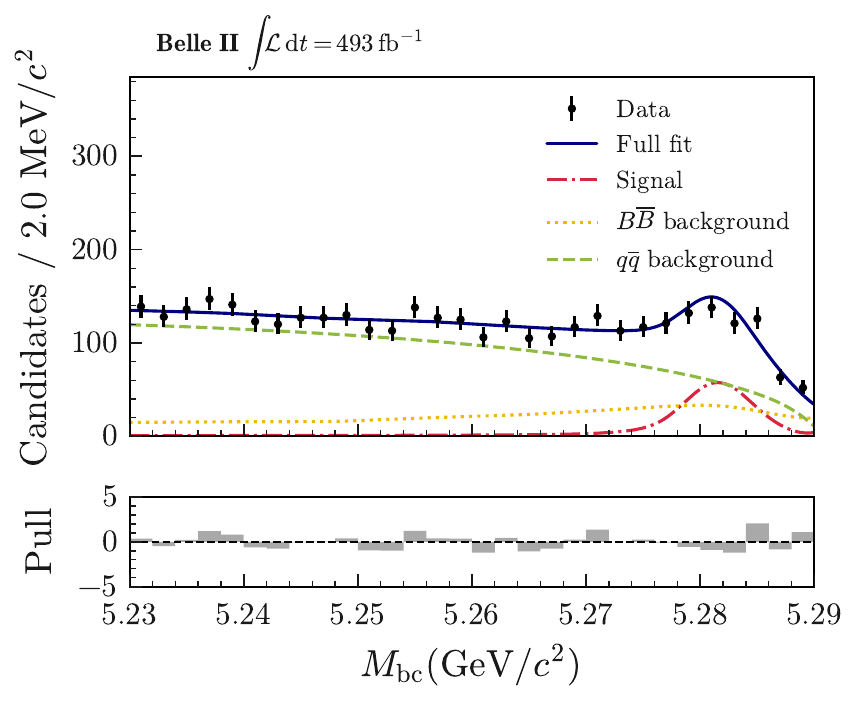}
      \end{subfigure}%
      \begin{subfigure}{.33\textwidth}
        \centering
        \includegraphics[width=.9\linewidth]{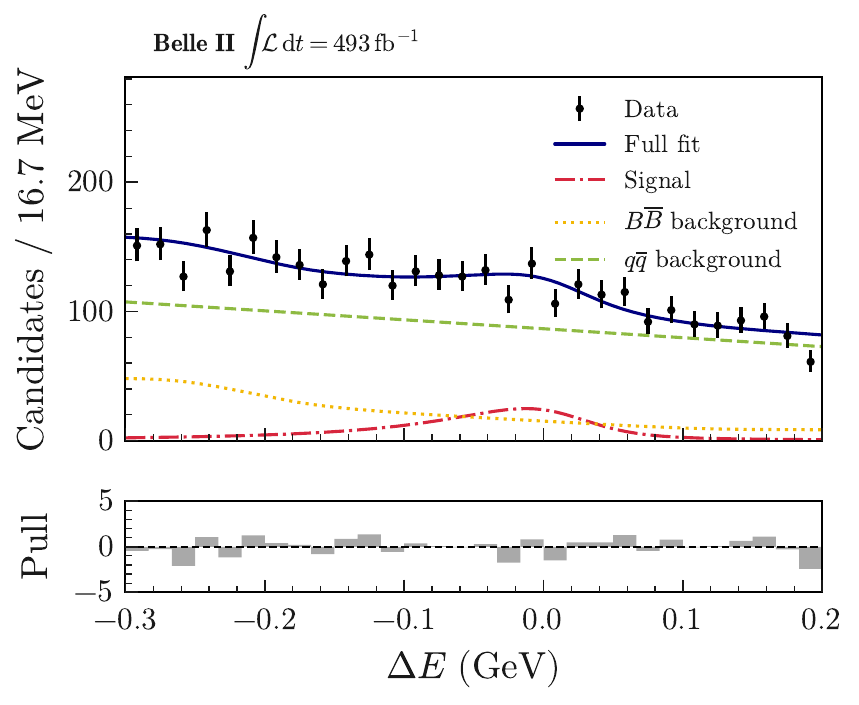}
      \end{subfigure}%
      \begin{subfigure}{.33\textwidth}
        \centering
        \includegraphics[width=.9\linewidth]{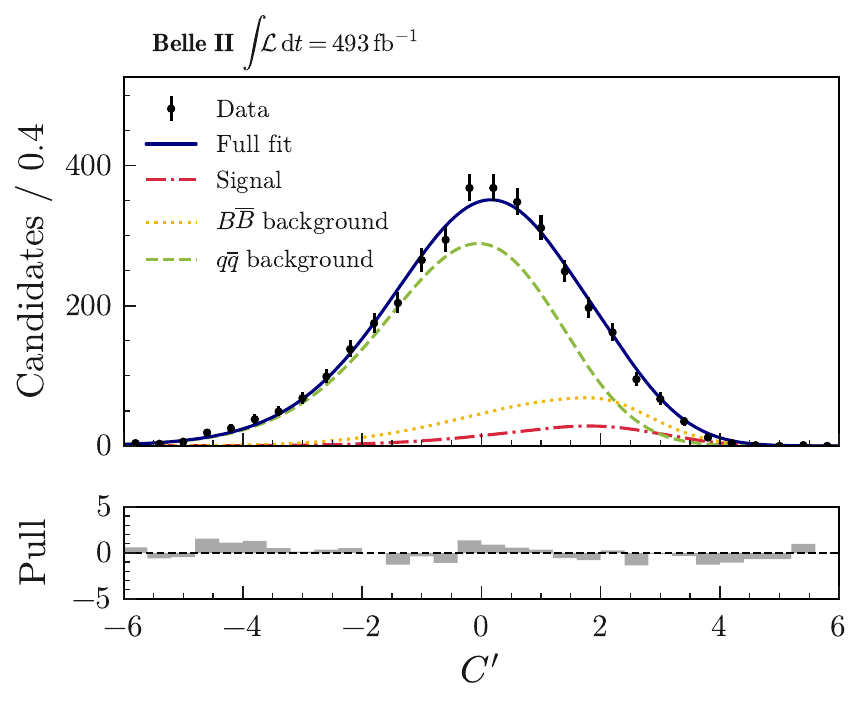}
      \end{subfigure}%
    \caption{\Mbc (left), \DeltaE (center), and $\CS$ (right) fit projections for MR1 (top) and MR2 (bottom) in Belle~II. Lower panels show the pulls, i.e., the differences between the data and fit results divided by the statistical uncertainty on the data.}
    \label{fig:se belle2}
\end{figure}

The projections of signal extraction PDFs (\Mbc, \DeltaE, and $\CS$) obtained from the full fit to data in the signal channel are shown in Figures~\ref{fig:se belle} and \ref{fig:se belle2} for Belle and Belle~II, respectively.

\subsection{\CP asymmetry PDF}

The \CP asymmetry PDF for signal comes into play in the fit performed to the \dt distribution to measure \SCP and \CCP. The physics-motivated model for \CP violation is modified to incorporate the effect of incorrect flavour assignment by the FT algorithm, and is convolved with a resolution function to account for finite detector resolution and various physical effects, such as secondary decays of charmed particles~\cite{PhysRevD.110.012001}:
\begin{multline}
    P_{\rm sig}(\Delta t, q \mid \sigma_{\Delta t}, r) = \frac{\exp(-|\Delta t|/\tau_{B})}{4\tau_{B}}
        \{1 -q\Delta w + q\mu(1-2w) \\
        +[q(1-2w) + \mu(1-q\Delta w)]\left[\SCP\sin(\dmd \Delta t) - \CCP\cos(\dmd \Delta t)\right]\} \otimes \mathcal{R}(\dt \mid \sigma_{\Delta t}).
\end{multline}
Here, $\tau_{B}$ is the lifetime of $B^0$ and $w$, $\Delta w$, and $\mu$ are the three FT parameters that represent the fraction of wrong tags, the asymmetry of wrong tags, and the asymmetry of the tagging efficiency, respectively. They are determined in seven bins of $r$ from a control channel of self-tagged $B^0 \to \jpsi K^*$ and $B^0 \to D^{(*)-} \pi^+$ decays for Belle and Belle~II, respectively~\cite{PhysRevD.110.012001}. The effective tagging efficiencies are ($33.6 \pm 1.2)\%$ and ($37.4 \pm 0.4)\%$ for Belle and Belle~II, respectively. We use linear interpolation to model the FT parameters as a function of $r$ and use a per-event dependence on $r$ in the fit.

The resolution function $\mathcal{R}_{\rm}(\dtres \equiv \dt - \dt_{\rm true} \mid \sigma_{\Delta t})$ is modelled as the sum of a core and a tail Gaussian ($G$) partially convolved with an exponential function, considering per-event dependence on $\sigma_{\Delta t}$, and is given by:
\begin{align}
    \mathcal{R}_{\rm{res}}(\dtres | \sdt) &= \brac{1-f_{\rm tail}} \mathcal{R}_{\rm{res}}^{\rm core}+ f_{\rm tail}\brac{\brac{1-f_{\rm exp}} \mathcal{R}_{\rm{res}}^{\rm tail-G} + f_{\rm exp} {\mathcal{R}_{\rm{res}}^{\rm tail-GE}}},\\
    \mathcal{R}_{\rm{res}}^{\rm core} &= G\brac{\dtres, \mu_{\rm core} \sdt, \sigma_{\rm core} \sdt},\\
    \mathcal{R}_{\rm{res}}^{\rm tail-G} &= G\brac{\dtres, \mu_{\rm tail} \sdt, \sigma_{\rm tail} \sdt},\\
    \mathcal{R}_{\rm{res}}^{\rm tail-GE} &= \mathcal{R}_{\rm{res}}^{\rm tail-G}\otimes\brac{f_{\rm R} \exp{\brac{\frac{-\dtres}{k\sdt}}}+ (1-f_{\rm R}) \exp{\brac{\frac{\dtres}{k\sdt}}}}.
\label{eqn:reso function}
\end{align}

The parameters of the resolution functions in data and simulation are determined from flavour-integrated fits to a control channel of $B^0 \to \jpsi\KS$ decays with the $\jpsi$ not included in the vertexing. This channel is a good proxy for the signal channel, with the only source of vertex information coming from the $\KS$ decay products.

The relative signal yields in the two flavours of $B_{\rm tag}$ contain information about the direct \CP violation parameter \CCP. In particular, the normalisation of the signal PDF for each event in the two flavours can be expressed in terms of \CCP as follows:
\begin{align}
    N_{\rm sig}^{q}(q=\pm1, r) = \frac{N_{\rm sig}}{2}\biggl(&1-q\Delta w + q\mu(1-2w) \nonumber \\
    &-\CCP\frac{q(1-2w)+\mu(1-q\Delta w)}{1+\dmd^2\tau_{\B}^2}\biggr).
    \label{eqn:CCP yields}
\end{align}

We model the \BBbar background \dt shape using the same functional form as the signal, but with effective lifetime ($\tau_{\BBbar}$) and \CP violation parameters ($S_{\BBbar}$ and $C_{\BBbar}$). We also use the same resolution function as that used for the signal, except for the width of the core Gaussian, which is determined from a fit to simulated \BBbar background events. The parameter $\tau_{\BBbar}$ is allowed to float in the full fit. On the other hand, $S_{\BBbar}$ and $C_{\BBbar}$ are fixed to their simulation values, determined as the weighted averages of the \CP violation parameters across the various background components. Furthermore, the yields of \BBbar background in the two $B_{\rm tag}$ flavours are parametrised in a similar way to the signal component, but using $C_{\BBbar}$ instead of \CCP. We model the \dt shape of the \qqbar component using the signal resolution function, with the mean and width of the core Gaussian allowed to float in the final fit. The \qqbar yield for each of the two $B_{\rm tag}$ flavours is half of the total continuum yield, as no \CP violation is expected in this component.

The \CP asymmetry fit strategy is validated using the control channel of $B^0 \to \jpsi \KS$ decays with the $\jpsi$ removed from the vertexing, with the same fit strategy as the signal channel. The fit quality is good with the \CP violation parameters measured as $S_{\jpsi \KS} = 0.65 \pm 0.06 \; (0.64 \pm 0.04)$ and $C_{\jpsi \KS} = 0.06 \pm 0.03 \; (-0.04 \pm 0.04)$ for Belle (Belle II), which are consistent with world averages~\cite{PDG2022}.

\subsection{Full fit}

The full fit consists of a simultaneous fit to TD and TI datasets in two flavours of $B_{\rm tag}$, with the respective PDFs as described above. The fit strategy is validated using large ensembles of pseudoexperiments generated from the PDFs (for  different input values of \SCP and \CCP), as well as samples drawn from MC simulations. The pulls are found to be normally distributed, and no significant bias is observed in the fitted \CP-violation parameters. Further validation is conducted by extracting the \B meson lifetime~\cite{PDG2022} and \CP violation parameters from the data using an ensemble of randomised flavour tags. In either case, the results are consistent with expectations.

\begin{figure}[!h]
    \centering
    \begin{subfigure}{.5\textwidth}
        \centering
        \includegraphics[width=.9\linewidth]{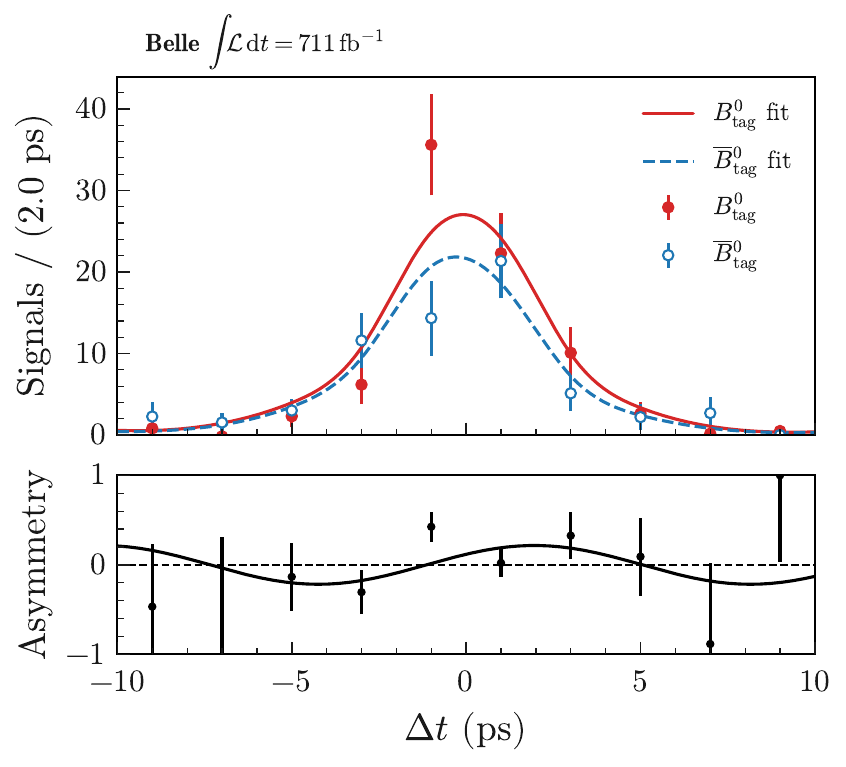}
      \end{subfigure}%
      \begin{subfigure}{.5\textwidth}
        \centering
        \includegraphics[width=.9\linewidth]{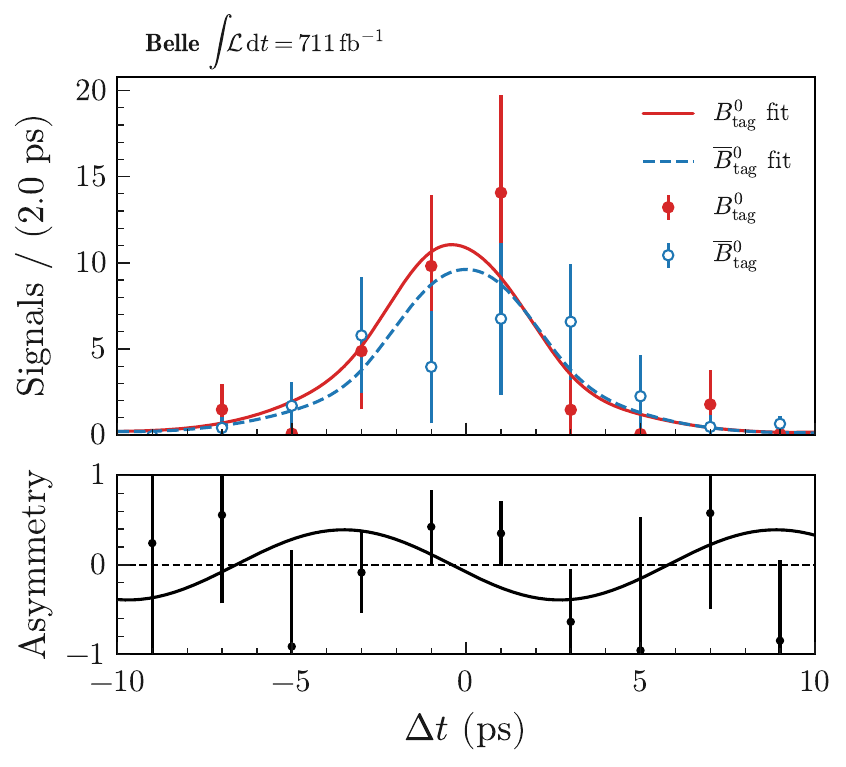}
      \end{subfigure}%
      \newline
      \begin{subfigure}{.5\textwidth}
        \centering
        \includegraphics[width=.9\linewidth]{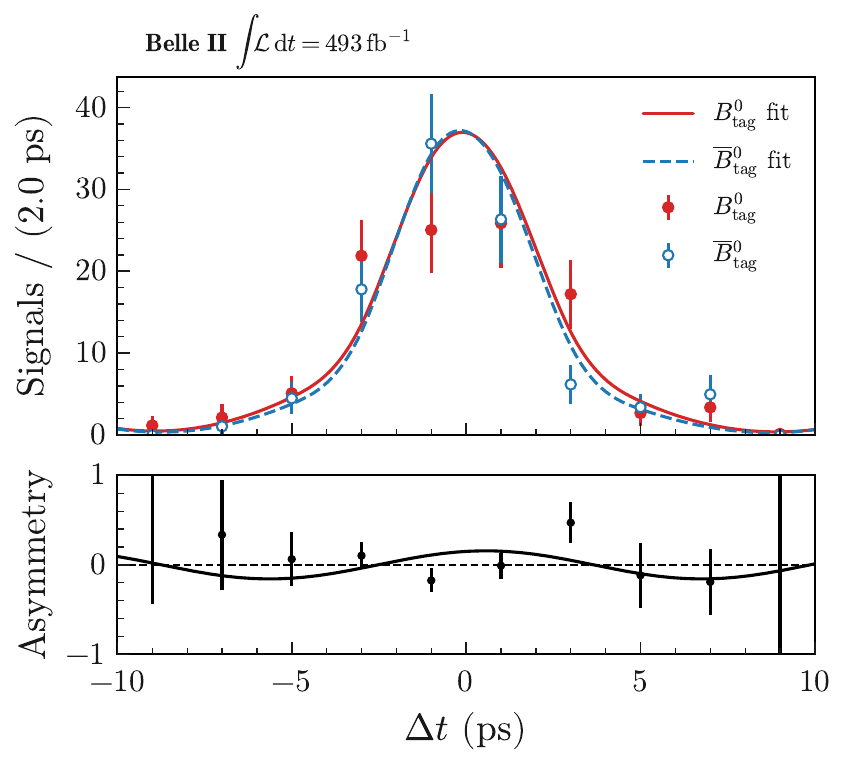}
      \end{subfigure}%
      \begin{subfigure}{.5\textwidth}
        \centering
        \includegraphics[width=.9\linewidth]{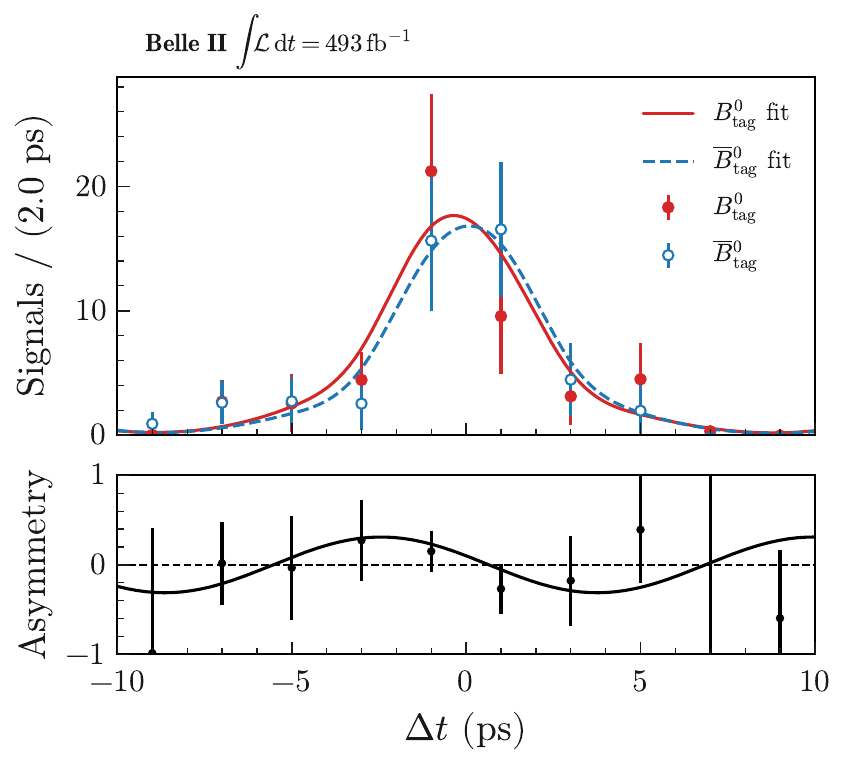}
      \end{subfigure}%
    \caption{Fit projections for background subtracted \dt distribution for the signal component in the TD dataset for MR1 (top left), MR2 (top right) in Belle, and MR1 (bottom left), MR2 (bottom right) in Belle~II. The candidates are weighted by $r$.}
    \label{fig:dt fit}
\end{figure}

The signal and background yields in the signal region, obtained from fits to the data, are given in Table~\ref{tab:signal_yields}. Similarly, the results for the \CP violation parameters are given in Table~\ref{tab:cpv_results}. The background-subtracted~\cite{Pivk:2004ty} \dt fit projections and asymmetries, $[N_{\rm sig}(\Bz_{\rm tag}) - N_{\rm sig}(\Bzb_{\rm tag})]/[N_{\rm sig}(\Bz_{\rm tag}) + N_{\rm sig}(\Bzb_{\rm tag})]$, for the signal component in the TD dataset are shown in Figure~\ref{fig:dt fit} for the two datasets and two mass regions. The \CCP information encoded in these projections corresponds to the TD dataset alone, which is consistent with that obtained from the TI dataset, within $1.4$ standard deviations; their combined result is presented in Table~\ref{tab:cpv_results}.

\begin{table}[h]
    \centering
    \caption{Signal and background yields in the signal region obtained from the fits to data. 
    The last column shows the ratio of the signal to background yield.}
    \label{tab:signal_yields}
    \begin{tabular}{llcccc}
        \hline
        Mass range & Dataset & $N_{\rm sig}$ & $N_{\BBbar}$ & $N_{\qqbar}$ & $N_{\rm sig}/(N_{\BBbar} + N_{\qqbar})$ \\
        \hline\hline 
        MR1 & Belle & $558 \pm 29$ & $30 \pm 9$ & $350 \pm 8$ & 1.47 \\
            & Belle~II & $452 \pm 24$ & $34 \pm 7$ & $178 \pm 6$ & 2.13 \\
        \hline
        MR2 & Belle & $212 \pm 26$ & $216 \pm 43$ & $\phantom{2}753 \pm 21$ & 0.22 \\
            & Belle~II & $213 \pm 22$ & $147 \pm 14$ & $\phantom{2}350 \pm 10$ & 0.43 \\
        \hline
    \end{tabular}
\end{table}

\begin{table}[h]
    \centering
    \caption{\CP violation parameters obtained from the full fits to data. The first uncertainties
listed are statistical, and the second are systematic; the latter are discussed in Section~\ref{systematics}.}
\begin{tabular}{llcc}
    \hline
    Mass range & Dataset & $\SCP$ & $\CCP$ \\
    \hline\hline 
    MR1 & Belle & $0.18 \pm 0.31 \pm 0.04$ & $-0.01 \pm 0.12 \pm 0.04$ \\
        & Belle~II & $0.04 \pm 0.19 \pm 0.02$ & $-0.16 \pm 0.10 \pm 0.04$ \\
        & \textbf{Belle + Belle~II} & $\mathbf{0.09 \pm 0.16 \pm 0.02}$ & $\mathbf{-0.09 \pm 0.08 \pm 0.04}$ \\
    \hline
    MR2 & Belle & $-0.37 \pm 0.58 \pm 0.09$ & $-0.01 \pm 0.29 \pm 0.07$ \\
        & Belle~II & $-0.29 \pm 0.41 \pm 0.09$ & $-0.11 \pm 0.22 \pm 0.09$ \\
        & \textbf{Belle + Belle~II} & $\mathbf{-0.32 \pm 0.33 \pm 0.09}$ & $\mathbf{-0.07 \pm 0.17 \pm 0.08}$ \\
    \hline
\end{tabular}
\label{tab:cpv_results}
\end{table}

The results for \SCP and \CCP are combined across the Belle and Belle~II datasets by calculating their weighted average, while accounting for correlations among the systematic uncertainties and observables~\cite{VALASSI2003391}. The correlations between the \SCP and \CCP values for the Belle\,+\,Belle~II dataset are approximately $+5\%$ in MR1 and $-7\%$ in MR2. The estimation of systematic uncertainties is discussed in Section~\ref{systematics}.

\subsection{Systematic uncertainties}
\label{systematics}
We consider several sources of systematic uncertainties that could affect our measurement of the \CP violation parameters. These are calculated separately for the Belle and Belle~II datasets in MR1 and MR2. The various sources of systematics and their contributions are summarised in Tables \ref{tab:total syst scp} and \ref{tab:total syst ccp} for \SCP and \CCP, respectively.

Systematic uncertainties due to the FT parameters ($w$, $\Delta w$, and $\mu$), physics parameters ($\tau_{B}$ and $\dmd$), \BBbar \CP asymmetries ($S_{\BBbar}$ and $C_{\BBbar}$), and the SCF fraction are estimated by sampling these parameters from Gaussian distributions and refitting the final dataset with these sampled values. The widths of the resulting distributions of fitted \SCP and \CCP values are then taken as the systematic uncertainties. For each FT parameter, the sampling Gaussian is centered at the mean value and assigned a width equal to the corresponding uncertainty, which arises from the limited size of calibration data samples. For physics parameters, the means and widths of the sampling Gaussians are set to their known values and uncertainties~\cite{PDG2022}. For \BBbar \CP asymmetries, the sampling Gaussian is centered at the values expected from MC simulation, with widths corresponding to the uncertainties in these parameters, typically around 10--20\%. For the SCF fraction, the sampling Gaussian is likewise centered at its MC expected value, with a width set to 20\% of that value.

The systematic uncertainties due to the fixed parameters of the resolution function and PDF shapes are estimated by sampling these parameters from the simulation-fit likelihood, and refitting the final dataset with these sampled values. The widths of the resulting distributions of fitted \CCP and \SCP values are taken as the systematic uncertainties. The systematic uncertainties due to non-parametric \BBbar KDE PDFs are estimated by generating new PDFs using datasets bootstrapped from simulated events. This method introduces Poisson fluctuations on the PDF shape. The final dataset is refitted using these generated KDE PDFs and the widths of the resulting distributions of fitted \CCP and \SCP values are taken as the systematic uncertainties.

Systematic uncertainties related to the $B$ meson vertex reconstruction stem from possible misalignment of vertex detectors, uncertainties in the IP profile, and corrections applied to charged track trajectories. These uncertainties are estimated using the same methodology as outlined in Refs.~\cite{PhysRevLett.108.171802, PhysRevD.110.012001}. The total vertexing systematic uncertainty is obtained by summing the individual contributions in quadrature. 

Systematic uncertainties are also calculated to account for the tag-side interference effect~\cite{PhysRevD.68.034010}. Additionally, the fit biases on \CP violation parameters obtained using randomised flavour tags are assigned as systematic uncertainties.

The uncertainties due to \BBbar \CP asymmetries, physics parameters, and tag-side interference are assumed to be 100\% correlated between Belle and Belle~II datasets.

\begin{table}[h]
    \centering
    \caption{Summary of systematic uncertainties on \SCP.}
    \begin{tabular}{l|cccc}
        \hline
        Source & Belle MR1 & Belle MR2 & Belle~II MR1 & Belle~II MR2 \\
        \hline\hline
        Flavour tagging & $\pm 0.010$ & $\pm 0.019$ & $\pm 0.002$ & $\pm 0.004$ \\
        Physics parameters & $\pm 0.001$ & $\pm 0.001$ & $\pm 0.001$ & $\pm 0.003$ \\
        \BBbar \CP asymmetries & $\pm 0.013$ & $\pm 0.076$ & $\pm 0.010$ & $\pm 0.081$ \\
        SCF fraction & $\pm 0.002$ & $\pm 0.006$ & $\pm 0.003$ & $\pm 0.005$ \\
        Resolution function & $\pm 0.018$ & $\pm 0.030$ & $\pm 0.007$ & $\pm 0.016$ \\
        PDF shape & $\pm 0.008$ & $\pm 0.024$ & $\pm 0.004$ & $\pm 0.030$ \\
        Vertex measurement & $\pm 0.023$ & $\pm 0.024$ & $\pm 0.005$ & $\pm 0.015$ \\
        Tag-side interference & $\pm 0.003$ & $\pm 0.004$ & $\pm 0.006$ & $\pm 0.001$ \\
        Fit bias & $\pm 0.005$ & $\pm 0.011$ & $\pm 0.001$ & $\pm 0.004$ \\
        \hline
        \textbf{Total} & \boldmath $\pm 0.035$ & \boldmath $\pm 0.091$ & \boldmath $\pm 0.016$ & \boldmath $\pm 0.090$ \\
        \hline
    \end{tabular}
    \label{tab:total syst scp}
\end{table}

\begin{table}[h]
    \centering
    \caption{Summary of systematic uncertainties on \CCP.}
    \begin{tabular}{l|cccc}
        \hline
        Source & Belle MR1 & Belle MR2 & Belle~II MR1 & Belle~II MR2 \\
        \hline\hline
        Flavour tagging & $\pm 0.021$ & $\pm 0.039$ & $\pm 0.007$ & $\pm 0.011$ \\
        Physics parameters & $\pm 0.001$ & $\pm 0.001$ & $\pm 0.001$ & $\pm 0.001$ \\
        \BBbar \CP asymmetries & $\pm 0.007$ & $\pm 0.046$ & $\pm 0.014$ & $\pm 0.084$ \\
        SCF fraction & $\pm 0.001$ & $\pm 0.001$ & $\pm 0.002$ & $\pm 0.002$ \\
        Resolution function & $\pm 0.002$ & $\pm 0.006$ & $\pm 0.003$ & $\pm 0.005$ \\
        PDF shape & $\pm 0.003$ & $\pm 0.013$ & $\pm 0.007$ & $\pm 0.011$ \\
        Vertex measurement & $\pm 0.006$ & $\pm 0.009$ & $\pm 0.002$ & $\pm 0.003$ \\
        Tag-side interference & $\pm 0.036$ & $\pm 0.036$ & $\pm 0.034$ & $\pm 0.036$ \\
        Fit bias & $\pm 0.001$ & $\pm 0.005$ & $\pm 0.003$ & $\pm 0.002$ \\
        \hline
        \textbf{Total} & \boldmath $\pm 0.043$ & \boldmath $\pm 0.072$ & \boldmath $\pm 0.038$ & \boldmath $\pm 0.093$ \\
        \hline
    \end{tabular}
    \label{tab:total syst ccp}
\end{table}

\section{Summary}
We report the measurement of time-dependent \CP violation parameters in \( \Bz \to \KS \piz \g \) decays using $770\times 10^6$ and $521\times 10^6$ \BBbar events from Belle and Belle~II, respectively. The combined values of the \CP violation parameters in the $K^{*0}(892)$ region ($M_{\KS\piz} \in [0.8,1.0]\gevcc$) are
$$\SCP = 0.09 \pm 0.16 \pm 0.02,$$
and
$$\CCP = -0.09 \pm 0.08 \pm 0.04.$$
Similarly, the combined values of the \CP violation parameters in the non-$K^{*0}(892)$ region ($M_{\KS\piz} \in (1.0,1.8]\gevcc$) are
$$\SCP = -0.32 \pm 0.33 \pm 0.09,$$
and
$$\CCP = -0.07 \pm 0.17 \pm 0.08.$$
These results represent the most precise measurements to date for \( \Bz \to \KS \piz \g \), improving upon the combination of the previous Belle~\cite{Belle:2006pxp} and Belle~II~\cite{PhysRevLett.134.011802} results by approximately 24\% for \SCP and 31\% for \CCP in the $K^{*0}(892)$ region, and supersede the earlier measurements~\cite{Belle:2006pxp,PhysRevLett.134.011802} for the same channel. These findings are consistent with the predictions of the Standard Model.

\section*{Acknowledgements}
This work, based on data collected using the Belle II detector, which was built and commissioned prior to March 2019,
and data collected using the Belle detector, which was operated until June 2010,
was supported by
Higher Education and Science Committee of the Republic of Armenia Grant No.~23LCG-1C011;
Australian Research Council and Research Grants
No.~DP200101792, 
No.~DP210101900, 
No.~DP210102831, 
No.~DE220100462, 
No.~LE210100098, 
and
No.~LE230100085; 
Austrian Federal Ministry of Education, Science and Research,
Austrian Science Fund (FWF) Grants
DOI:~10.55776/P34529,
DOI:~10.55776/J4731,
DOI:~10.55776/J4625,
DOI:~10.55776/M3153,
and
DOI:~10.55776/PAT1836324,
and
Horizon 2020 ERC Starting Grant No.~947006 ``InterLeptons'';
Natural Sciences and Engineering Research Council of Canada, Digital Research Alliance of Canada, and Canada Foundation for Innovation;
National Key R\&D Program of China under Contract No.~2024YFA1610503,
and
No.~2024YFA1610504
National Natural Science Foundation of China and Research Grants
No.~11575017,
No.~11761141009,
No.~11705209,
No.~11975076,
No.~12135005,
No.~12150004,
No.~12161141008,
No.~12405099,
No.~12475093,
and
No.~12175041,
and Shandong Provincial Natural Science Foundation Project~ZR2022JQ02;
the Czech Science Foundation Grant No. 22-18469S,  Regional funds of EU/MEYS: OPJAK
FORTE CZ.02.01.01/00/22\_008/0004632 
and
Charles University Grant Agency project No. 246122;
European Research Council, Seventh Framework PIEF-GA-2013-622527,
Horizon 2020 ERC-Advanced Grants No.~267104 and No.~884719,
Horizon 2020 ERC-Consolidator Grant No.~819127,
Horizon 2020 Marie Sklodowska-Curie Grant Agreement No.~700525 ``NIOBE''
and
No.~101026516,
and
Horizon Europe Marie Sklodowska-Curie Staff Exchange project JENNIFER3 Grant Agreement No.~101183137 (European grants);
L’Institut National de Physique Nucl\'eaire et de Physique des
Particules (IN2P3) du CNRS under Project Identification No.
CNRS-IN2P3-14-PP-033
and L’Agence Nationale de la Recherche (ANR) under Grant No. ANR-23-CE31-
0018 and ANR-25-CE31-1333 (France);
BMFTR, DFG, HGF, MPG, and AvH Foundation (Germany);
Department of Atomic Energy under Project Identification No.~RTI 4002,
Department of Science and Technology,
and
UPES SEED funding programs
No.~UPES/R\&D-SEED-INFRA/17052023/01 and
No.~UPES/R\&D-SOE/20062022/06 (India);
Israel Science Foundation Grant No.~2476/17,
U.S.-Israel Binational Science Foundation Grant No.~2016113, and
Israel Ministry of Science Grant No.~3-16543;
Istituto Nazionale di Fisica Nucleare and the Research Grants BELLE2,
and
the ICSC – Centro Nazionale di Ricerca in High Performance Computing, Big Data and Quantum Computing, funded by European Union – NextGenerationEU;
Japan Society for the Promotion of Science, Grant-in-Aid for Scientific Research Grants
No.~16H03993,
No.~16H06492,
No.~16K05323,
No.~17H01133,
No.~17H05405,
No.~18K03621,
No.~18H03710,
No.~18H05226,
No.~19H00682, 
No.~20H05850,
No.~20H05858,
No.~22H00144,
No.~22K14056,
No.~22K21347,
No.~23H05433,
No.~26220706,
No.~26400255,
and
No.~26H02056,
and
the Ministry of Education, Culture, Sports, Science, and Technology (MEXT) of Japan;  
National Research Foundation (NRF) of Korea Grants
No.~2021R1-F1A-1064008,
No.~2022R1-A2C-1003993,
No.~RS-2018-NR031074,
No.~RS-2021-NR060129,
No.~RS-2024-00354342,
No.~RS-2025-02219521,
No.~RS-2026-25471491,
No.~RS-2026-25480677,
and
No.~RS-2026-25486791,
Radiation Science Research Institute,
Foreign Large-Size Research Facility Application Supporting project,
the Global Science Experimental Data Hub Center, the Korea Institute of Science and
Technology Information (K26L1M2C3)
and
KREONET/GLORIAD;
Universiti Malaya RU grant, Akademi Sains Malaysia, and Ministry of Education Malaysia;
Frontiers of Science Program Contracts
No.~FOINS-296,
No.~CB-221329,
No.~CB-236394,
No.~CB-254409,
and
No.~CB-180023, and SEP-CINVESTAV Research Grant No.~237 (Mexico);
the Polish Ministry of Science and Higher Education and the National Science Center;
the Ministry of Science and Higher Education of the Russian Federation
and
the HSE University Basic Research Program, Moscow;
University of Tabuk Research Grants
No.~S-0256-1438 and No.~S-0280-1439 (Saudi Arabia);
Slovenian Research Agency and Research Grants
No.~J1-50010
and
No.~P1-0135;
Ikerbasque, Basque Foundation for Science,
State Agency for Research of the Spanish Ministry of Science and Innovation through Grant No. PID2022-136510NB-C33, Spain,
the Severo Ochoa project CEX2023-001292-S funded by MICIU/AEI, State Secretariat for
Telecommunications and Digital Infrastructure with reference
TSI-069100-2023-0012, State Agency for Research of the Spanish Ministry
of Science, Innovation and Universities through Grant No
PID2024-156645NB-C21;
the Swiss National Science Foundation;
The Knut and Alice Wallenberg Foundation (Sweden), Contracts No.~2021.0174, No.~2021.0299, and No.~2023.0315;
National Science and Technology Council,
and
Ministry of Education (Taiwan);
Thailand Center of Excellence in Physics;
TUBITAK ULAKBIM (Turkey);
National Research Foundation of Ukraine, Project No.~2020.02/0257,
and
Ministry of Education and Science of Ukraine;
the U.S. National Science Foundation and Research Grants
No.~PHY-1913789 
and
No.~PHY-2111604, 
and the U.S. Department of Energy and Research Awards
No.~DE-AC06-76RLO1830, 
No.~DE-SC0007983, 
No.~DE-SC0009824, 
No.~DE-SC0009973, 
No.~DE-SC0010007, 
No.~DE-SC0010073, 
No.~DE-SC0010118, 
No.~DE-SC0010504, 
No.~DE-SC0011784, 
No.~DE-SC0012704, 
No.~DE-SC0019230, 
No.~DE-SC0021616, 
No.~DE-SC0022350, 
No.~DE-SC0023470; 
and
the Vietnam Academy of Science and Technology (VAST) under Grant
No.~DL0000.05/26-27.

These acknowledgements are not to be interpreted as an endorsement of any statement made
by any of our institutes, funding agencies, governments, or their representatives.

We thank the SuperKEKB team for delivering high-luminosity collisions;
the KEK cryogenics group for the efficient operation of the detector solenoid magnet and IBBelle on site;
the KEK Computing Research Center for on-site computing support; the NII for SINET6 network support;
and the raw-data centers hosted by BNL, DESY, GridKa, IN2P3, INFN, 
PNNL/EMSL, 
and the University of Victoria.

\ifdefvoid{\hepdata}{}{Numerical data corresponding to the results presented are available as HEPData.}
The full Belle II data are not publicly available.
The collaboration will consider requests for access to the data that support this article.

\bibliographystyle{JHEP}
\bibliography{belle2-references,references}

\end{document}